\documentclass[12pt]{iopart}
\usepackage{graphicx}
\usepackage{amsfonts}
\usepackage{amssymb}
\usepackage{color}
\usepackage{url}

\def\bra#1#2{\left#1\rule{0ex}{#2ex}\right.}

\begin{document}

\title[]{Fano-like interference of plasmon resonances at a single rod-shaped nanoantenna}

\author{F L\'opez-Tejeira, R Paniagua-Dom\'{\i}nguez, \newline
R Rodr\'{\i}guez-Oliveros and J A S\'anchez-Gil}

\address{Instituto de Estructura de la Materia (IEM-CSIC), Consejo Superior de Investigaciones Cient\'{\i}ficas, Serrano 121, E-28006 Madrid, Spain}
\ead{flt@iem.cfmac.csic.es}

\begin{abstract}
Single metallic nanorods acting as half-wave antennas in the optical range exhibit an asymmetric, multi-resonant scattering spectrum  that strongly depends on both their length and dielectric properties. Here we show that such spectral features can be easily understood in terms of Fano-like interference between adjacent plasmon resonances. On the basis of analytical and numerical results for different geometries, we demonstrate that Fano resonances may appear for such single-particle nanoantennas provided that interacting resonances overlap in both spatial and frequency domains.
\end{abstract}
\maketitle

\section{Introduction}
\label{intro}
Different experimental \cite{Muhlschlegel2005}-\cite{Neubrech2008} and purely theoretical \cite{Aizpurua2005}-\cite{Alu2008} investigations have shown that metallic nanorods act as standing-wave resonators for localized plasmon resonances in the optical regime, thus exhibiting geometrical half-wavelength resonances with spectral positions depending mainly on the length of the rods. This particular type of so-called ``optical nanoantennas'' have raised the prospect of significant improvements in fields such as photodetection \cite{Knight2011}, field-enhanced spectroscopy \cite{Ming2009} or control of emission direction in single-molecule light sources \cite{Taminiau2008}.

Generally speaking, most of device-oriented studies are focused on nanoantennas operating at the dipole-like resonance. However, structures with a high aspect ratio may support additional resonances that have usually been the subject of a more fundamental research work. Hence, several authors have already elucidated the scaling properties of high-order longitudinal modes, as well as their dependence on shape, size, orientation and dielectric environment by means of diverse approaches and techniques \cite{Payne2006,Khlebtsov2007}, \cite{Ghenuche2008}-\cite{Wei2010}. Nevertheless, a relevant issue has yet to be addressed for multi-resonant nanoantennas, that is the emergence of asymmetric line profiles in single-particle extinction or scattering spectra. Interestingly, such a feature seems to go  almost unnoticed for the nanoplasmonics community, despite being apparent in some previous references \cite{Payne2006,Khlebtsov2007,Chau2009,Wei2010}.  In fact, to the best of our knowledge,  the only explicit report on the occurrence of Fano-like asymmetric line shapes in the scattering spectra of a single silver nanorod can be found in a recent paper by Reed {\it et al.} \cite{Reed2011}, though the emphasis is put on dimer structures therein.

In this work, we show that these asymmetric line profiles can be easily understood in terms of the so-called Fano-like interference between localized plasmon resonances that has been recently reported for a variety of coupled metal nanoparticles \cite{Miroshnichenko2010}-\cite{Gallinet2011}. Being more precise, we present a simplified analytical model that describes spectral features of a single rod-shaped nanoantenna in terms of Fano-like interference. Contrary to the common assumption that interference does not play any role in total scattering or extinction of a single metallic surface, we find a good agreement with numerical results, which are attained through the separation of variables (SVM), finite element (FEM) and surface integral equation (SIEM) methods (see \ref{calcu} for a succinct description of calculation techniques). Furthermore, we make use of explicit expressions for light scattering by spheroids to conclude that not only spectral but also spatial overlap (i.e. non-orthogonality) between interacting modes underlies the emergence of such single-rod resonances. This points out the need of being extremely cautious when applying the premises of standard Mie theory to particles that significantly depart from sphericity.

\section{Single metallic nanorods acting as half-wave nanoantennas}
\subsection{Fano-like interference of longitudinal plasmon resonances}
Let us begin by briefly reviewing the basics of light scattering by a single metallic nanorod with dielectric function $\varepsilon_m$ that is surrounded by a medium with constant permittivity $\varepsilon_d$: On the assumption that rod diameter $D$ is much smaller than its total length $L$, the electromagnetic response to $p$-polarized light impinging perpendicular to the long side is fully governed by longitudinal modes. The fundamental resonance $\lambda_{res}^{(1)}$ can thus be described as a dipolar excitation of charges at the rod surface, with its wavelength exhibiting a linear dependence on $L$ , $\lambda_{res}^{(1)} \propto 2L$. For a perfectly conducting material of negligible thickness, $\lambda_{res}^{(1)}$ is precisely equal to $2L$, whereas $\lambda_{res}^{(1)}\gg 2L$ at optical frequencies \cite{Khlebtsov2007,Novotny2007,Bryant2008}. As rod length increases, additional resonances may appear. Following Khlebtsov and Khlebtsov \cite{Khlebtsov2007}, we assume the position of any longitudinal resonance to be described by the following approximate scaling law
\begin{equation}
\lambda_{res}^{(n)} \approx B_0 +B_1 \frac{L}{nD}
\label{scaling}
\end{equation}
where $n$ is an odd integer and  $B_0, B_1$ are coefficients that depend on the actual geometry and dielectric environment of the system and have to be determined from linear regression. With respect to $B_0, B_1$, we also have to notice that, although not used in this paper, explicit expressions can be obtained within the framework of Novotny's model for effective wavelength scaling at optical antennas \cite{Novotny2007}.

As mentioned in the previous section, the interaction of adjacent resonances has been found to be compatible with a Fano-like interference model \cite{Ropers2005},\cite{Lukyanchuk2010}-\cite{Fan2010}, where the lower resonance plays the role of continuum in canonical Fano line shape \cite{Fano1961}. Given that light scattering by a rod of finite length cannot be treated in a closed form within the framework of standard Mie theory \cite{BHBook}, we assume a heuristic line shape for the first approach. Hence,
\begin{equation}
Q_{sca} \approx |f(\omega)|^2\label{lineshape}
\end{equation}
with
\begin{equation}
f(\omega)\equiv A(\omega)+F\bra{[}{4}\frac{b_1}{(\omega-\omega_1)+i b_1}+\frac{q b_3}{(\omega-\omega_3)+i b_3}\bra{]}{4}\label{lineshape2}
\end{equation}
where $A(\omega)$ is a slowly varying amplitude and  $F$ stands for the complex amplitude of the fundamental resonance, which is described in terms of its real central frequency $\omega_1$ and spectral width $b_1$. Dimensionless real parameter $q$ modulates the interaction with the adjacent plasmon resonance (analogously defined by $\omega_3$ and $b_3$), thus governing the line shape asymmetry.
\begin{figure}
  \flushright
  \includegraphics[width=0.8\textwidth]{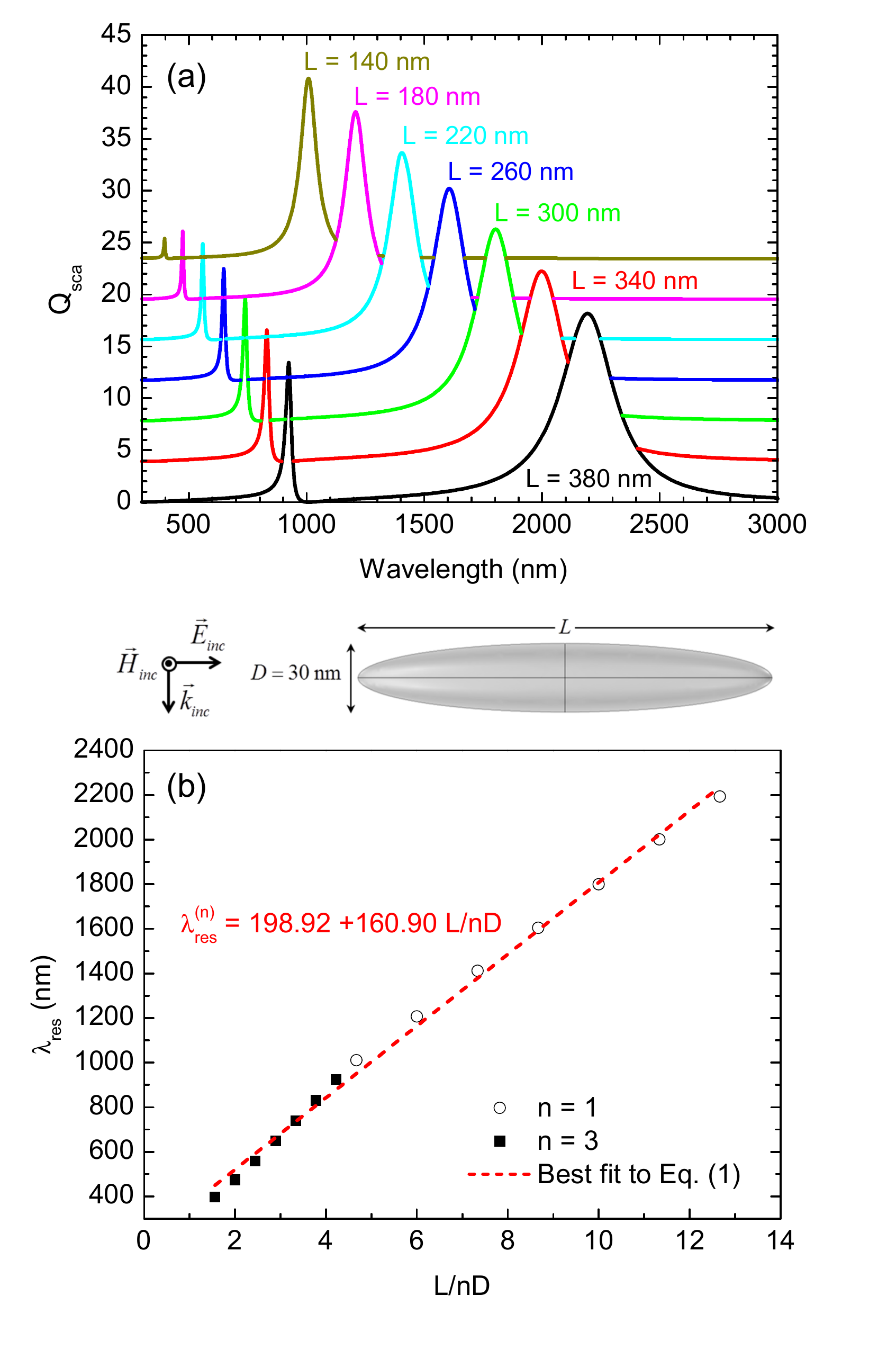}\\
  \caption{(a) Calculated scattering efficiency as a function of wavelength for a single Ag spheroid surrounded by glass ($\varepsilon_d=2.25$). Incident field is $p$-polarized and impinges perpendicular to the rotation axis of the spheroid. Different curves correspond to increasing values of $L$, whereas $D$ is set to 30 nm for all calculations. (b) Linear scaling of resonant wavelengths in panel (a) as a function of normalized aspect ratio $L/nD$. Dashed line marks the best fit to Eq. \eref{scaling}.}\label{Qscavsvlan}
\end{figure}

\subsection{Prolate spheroidal nanorods}
From the computational point of view, the shape of half-wave nanoantennas is usually modeled by a right circular cylinder with either flat or hemispherical ends (spherocylinder). However, we have found the prolate spheroid to be the most convenient geometry to start with, because of the following reasons: (i) Previous works \cite{Asano1975,Voshchinnikov1993} on the basis of SVM have provided us with a very efficient approach to calculate extinction, scattering and absorption cross-sections even for very elongated nanoantennas; (ii) In addition to its low numerical cost, such a theoretical framework also makes apparent the origin of unexpected asymmetry in line profiles, as detailed hereafter; (iii) Prolate spheroids are not only of academic interest, as they accurately describe the so-called nanorice structures \cite{Wei2010,Wang2006}.

In \fref{Qscavsvlan} we present the calculated scattering efficiency $Q_{sca}$ for a single silver spheroid surrounded by glass ($\varepsilon_d=2.25$) under the assumption that incident field is $p$-polarized and impinges perpendicular to the long side of the rod. Different curves correspond to increasing values of total length $L$ within the $[100,400]$ nm range, whereas the polar diameter $D$ is set to 30 nm for all calculations. As can be seen, the position of resonances increases linearly with $L$ and  such displacement is fairly well described by \eref{scaling} (see \fref{Qscavsvlan}(b)). For $L/D \gtrsim 5$, the peaks arising from resonances with $n=1$ and $n=3$ are clearly apparent, as it is the asymmetry of the line shape between them.

\begin{figure}
 \flushright
 \includegraphics[width=0.8\textwidth]{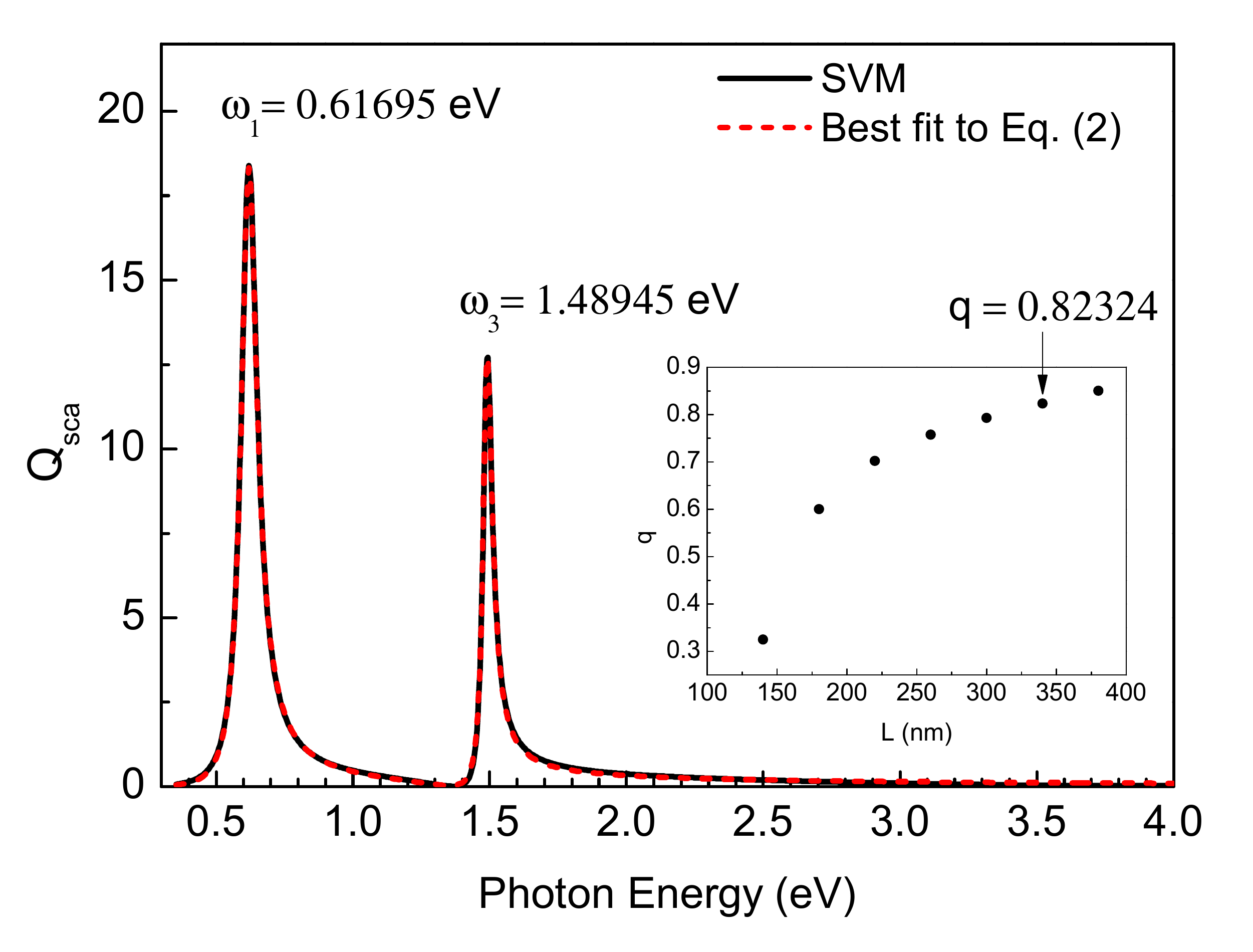}\\
  \caption{Calculated scattering efficiency (solid line) as a function of photon energy for a single Ag spheroid ($L=340$ nm; $D= 30$ nm) in $\varepsilon_d=2.25$. Incident field is $p$-polarized and impinges perpendicular to the rotation axis of the spheroid. Dashed line renders the best-fitting curve to Eq. \eref{lineshape}. Obtained values of $\omega_1,\omega_3$ and $q$ are also shown. Lower right inset panel depicts the obtained values of $q$ for every $L$ in \fref{Qscavsvlan}.}\label{QscavseV}
\end{figure}
In order to test out our approach, let us take a closer look to the curve corresponding to 340 nm-long spheroid. \Fref{QscavseV} renders the calculated scattering efficiency as a function of energy and its best-fitting curve to \eref{lineshape}. It may be seen that our heuristic Fano-like line shape agrees very well with the full electromagnetic calculation. Besides, the obtained value $q=0.82324$ is consistent with asymmetric profiles being described by $|q| \approx 1$ \cite{Miroshnichenko2010}. Such an agreement extends to the whole range of rod lengths in \fref{Qscavsvlan}, as summarized in the inset panel (See \ref{fitt2fano}).

 \begin{figure}
  \flushright
  \includegraphics[width=0.8\textwidth]{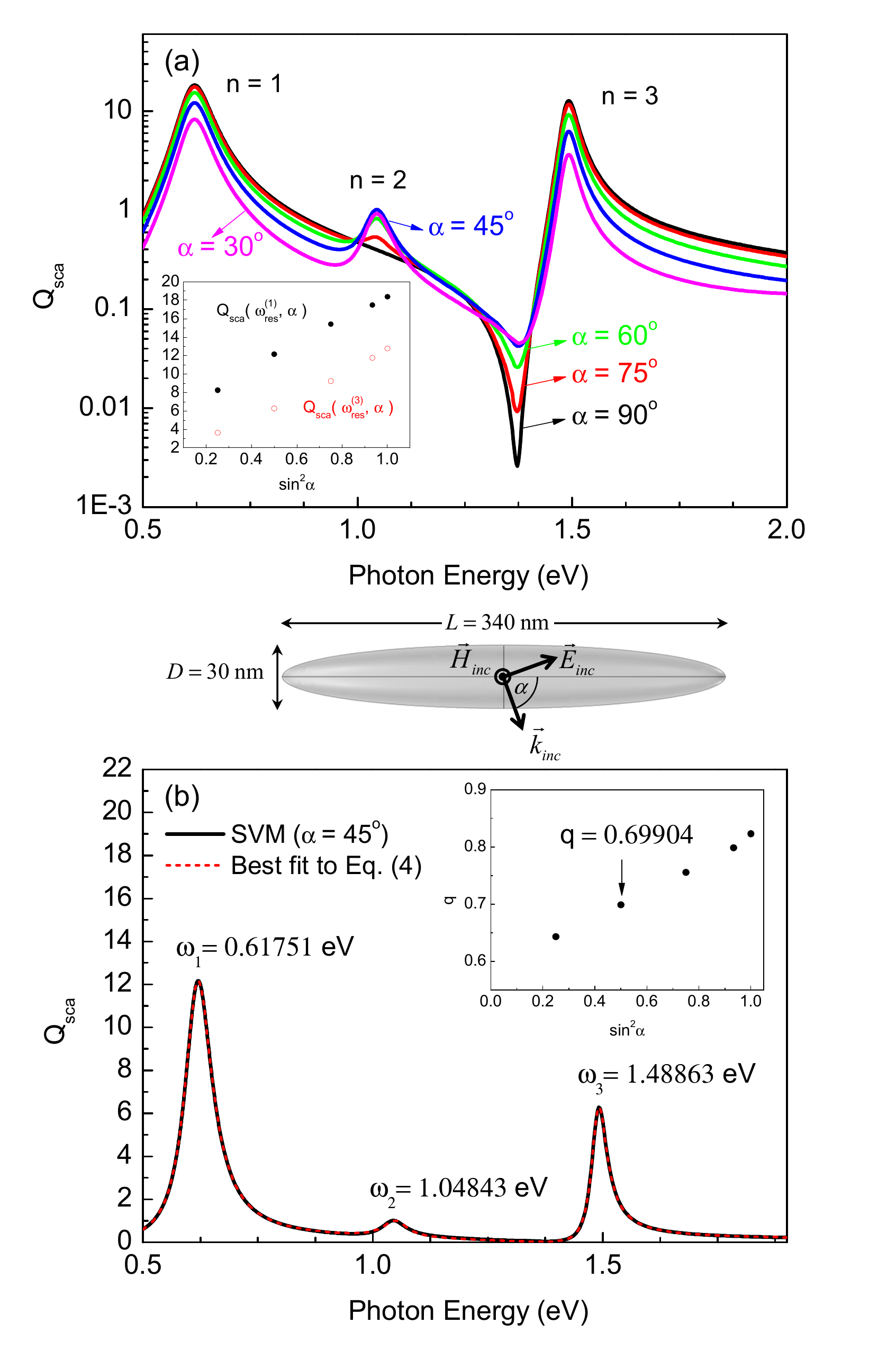}\\
  \caption{(a) Calculated $Q_{sca}$ as a function of photon energy for a single Ag spheroid ($L=340$ nm; $D= 30$ nm) surrounded by $\varepsilon_d=2.25$. Incident field is $p$-polarized and impinges with angle $\alpha$ with respect to the rotation axis. Different curves correspond to decreasing values of $\alpha$. The intensities of ``odd'' peaks as a function of $\alpha$ are presented at lower left inset panel. (b) Comparison between calculated $Q_{sca}$ and its best-fitting curve to Eq. \eref{lineshape3} for $\alpha=45^\mathrm{o}$. Obtained values of $\omega_1,\omega_2,\omega_3$ and $q$ are also shown. Upper right inset panel depicts the obtained values of $q$ for every $\alpha$ in (a).}\label{obliki}
\end{figure}
 For the case of oblique incidence (i.e. if the angle $\alpha$ between the incident light vector $\mathbf{k}_{inc}$ and the rotation axis of the spheroid is not equal to $90^\mathrm{o}$), ``even'' modes become accessible \cite{Payne2006,Khlebtsov2007, Wei2010} and should therefore be incorporated to our analysis. Different curves in \ref{obliki}(a) correspond to the calculated $Q_{sca}$ for decreasing values of $\alpha$ and the same geometry, polarization and dielectric environment as in \fref{QscavseV}. Logarithmic scale is used for optimal visualization of less intense features. As can be seen, the oblique line shapes differ quite little from that for normal incidence at the vicinity of resonances with labels $n=1,3$, except from that the intensities of ``odd'' peaks exhibit a linear dependence with $\sin^2 \alpha$ (see lower left inset panel). However, an extra resonance develops just on top of base line at $\omega \approx 1.045$ eV. Given that it is located within the $[\omega_1,\omega_3]$ range, we label it as $n=2$. This new peak reaches its maximum for $\alpha=45^\mathrm{o}$ and seems to be  symmetrical with respect to its central frequency. Hence, we can conjecture that symmetry precludes interference between longitudinal modes with different parity, so that only an additive term accounts for the contribution of this resonance  to $Q_{sca}$,
\begin{equation}
Q_{sca}(\omega,\alpha \neq 90^\mathrm{o}) \approx |f(\omega)|^2+\frac{|F_2|^2 b_2^2}{b_2^2+(\omega-\omega_2)^2}\label{lineshape3}
\end{equation}
\noindent where $F_2,\omega_2,b_2 $ are the complex amplitude, the central frequency, and the spectral width of the resonance with label $n=2$, respectively.

Comparison between calculated $Q_{sca}(\omega,\alpha=45^\mathrm{o})$ and its best-fitting curve to \eref{lineshape3} in \fref{obliki}(b) confirms our hypothesis of the lack of interaction between ``odd'' and ``even'' resonances. As summarized in the upper right inset panel, the only effect of oblique incidence (aside from the emergence of a new peak!) is the already mentioned global quenching of $Q_{sca}(\omega_{res}^{(1),(3)})$ (i.e.,the intensities of ``odd'' peaks) and therefore that of $q$. Being the understanding of line shape asymmetry the main goal of our work, we do not discuss on oblique incidence any further, except for the detailed description of the dependence of different parameters on $\alpha$ that is presented in \ref{fitt2fano}. However, we have to point out that inter-parity coupling may be allowed for a symmetry-broken configuration, such as depositing the nanospheroid onto a dielectric substrate, as recently proposed for the nanocube geometry \cite{Zhang2011}. In fact, we can easily envisage that the two different intra- and inter-parity mechanisms' operating simultaneously opens a very interesting scenario that certainly warrants further investigation.

Going beyond heuristic description brings us up against the actual meaning of \eref{lineshape2}. From a formal point of view, it is clear that $f(\omega)$ plays the role of effective polarizability in a somehow generalized Rayleigh-Gans theory for nonspherical particles. Nevertheless, there is still the concern of how to explain the emergence of asymmetry in line profiles. As previously mentioned, Fano resonances require an observable that is sensitive to interference, but standard Mie theory predicts the total scattering by a single metallic surface to be proportional to the mere sum of intensities.

Such a discrepancy can be easily explained for the case of prolate spheroidal particles by means of the SVM formalism. According to Reference \cite{Voshchinnikov1993}, the scattering efficiency of a prolate spheroid for $p$-polarized light impinging at normal incidence is given by
\begin{eqnarray}
\fl Q_{sca}=\frac{4}{L D \,k_d^2} \bra{\{}{4} 2\sum_{l=1}^{\infty}|b_l^{(d)}|^2 N_{1l}^2 (c_d)+\mathrm{Re} \sum_{l=1}^{\infty}\sum_{m=l}^{\infty}\sum_{n=m}^{\infty}\mathit{i}^{\ n-l}\bra{[}{3}k_d^2\, a_{ml}^{(d)}\Big(a_{mn}^{(d)}\Big)^*\omega_{ln}^{(m)}(c_d,c_d) +{}\nonumber\\
+\ \mathit{i} k_d \bra{(}{3}b_{ml}^{(d)}\Big(a_{mn}^{(d)}\Big)^*\kappa_{ln}^{(m)}(c_d,c_d)-a_{ml}^{(d)}\Big(b_{mn}^{(d)}\Big)^*\kappa_{nl}^{(m)}(c_d,c_d)\bra{)}{3}+{} \label{Qscaspher}\\
+\ b_{ml}^{(d)}\Big(b_{mn}^{(d)}\Big)^*\tau_{ln}^{(m)}(c_d,c_d)\bra{]}{3}N_{ml}(c_d)N_{mn}(c_d)\bra{\}}{4}\nonumber
\end{eqnarray}

Without entering into further details, let us say that $\kappa_{ln}^{(m)}$, $\tau_{ln}^{(m)}$ and $\omega_{ln}^{(m)}$ are different integrals of the normalized prolate angular spheroidal wave functions with normalization coefficients $N_{ij}(c_d)$, where $c_d=k_d/2\ \sqrt{L^2-D^2}$, $k_d=\sqrt{\varepsilon_d}\ 2 \pi/ \lambda$ and $\lambda$ is the wavelength of incoming (or outgoing) radiation. More importantly, $a_{ij}^{(d)}$ and $b_l^{(d)}$, $b_{ij}^{(d)}$ represent the expansion coefficients of the azimuthal components of scattered electric and magnetic fields respectively. Single or double subscript marks whether the expansion accounts for the axisymmetric (i.e. the one that does not depend on the azimuthal angle) or non-axisymmetric (i.e. azimuth-dependent) part of the corresponding field. For the axisymmetric part, the introduction of Debye's potential representation of vector fields (i.e. the potentials that solve the light scattering problem for spheres) leads to the term following the curly bracket in \eref{Qscaspher}, which closely resembles Mie's result. However, a proper representation of the non-axisymmetric part of the electromagnetic fields requires a combination of Debye's potential with that used to deal with
light scattering by an infinitely long cylinder, namely Hertz's \cite{Voshchinnikov2000}. Hence double-subscript interference terms arise in $Q_{sca}$, thus providing a mathematical description for the interaction between adjacent resonances in prolate spheroids. Note that such interference terms are not negligible only if the interacting fields are non-orthogonal, that is, spatial overlap between interfering plasmon modes is required for Fano-like interference.

\begin{figure}
  \flushright
  \includegraphics[width=0.84\textwidth]{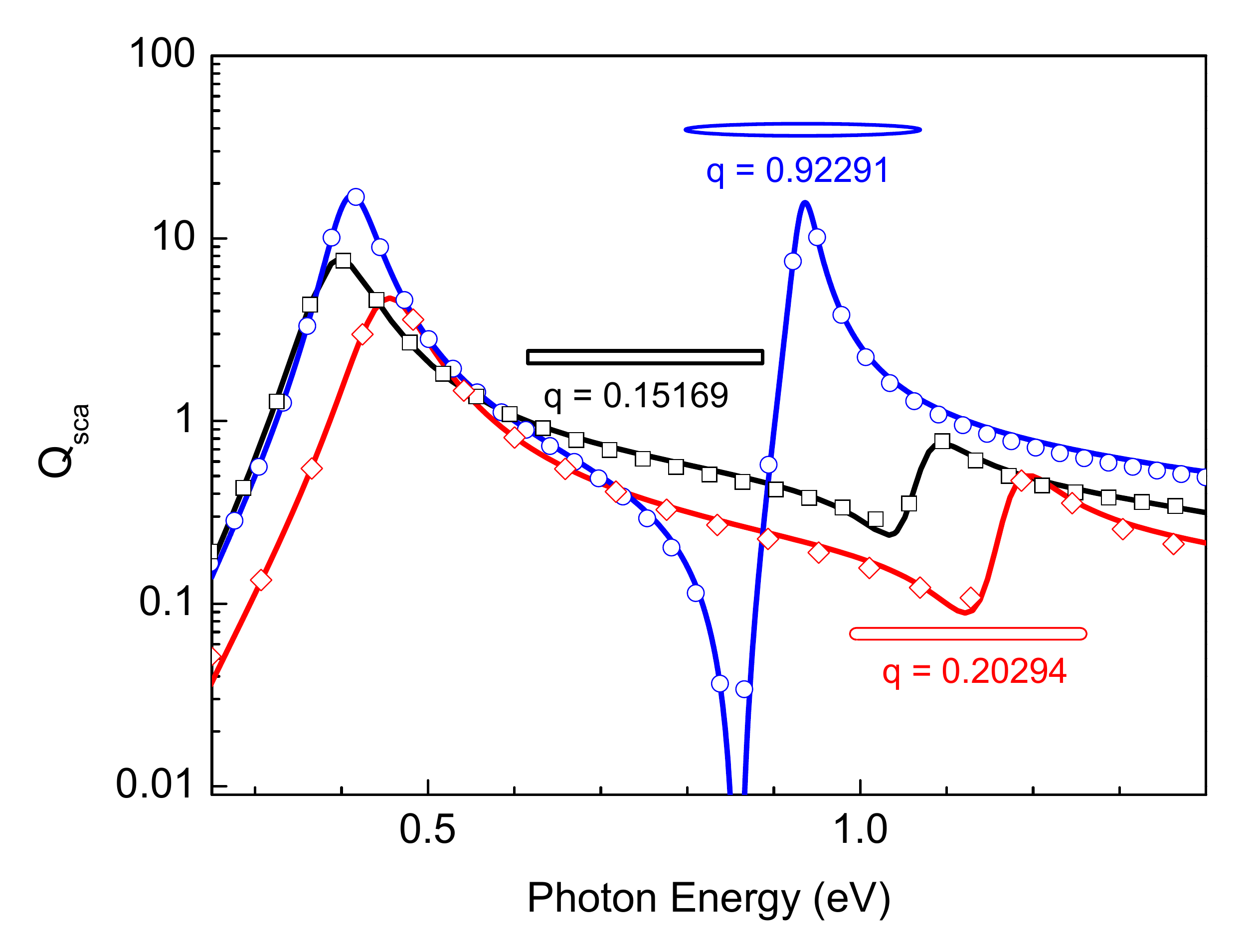}\\
  \caption{Calculated scattering efficiency as a function of photon energy for a single Ag nanorod ($L=550$ nm; $D= 30$ nm) surrounded by glass ($\varepsilon_d=2.25$). Incident field is $p$-polarized and impinges perpendicular to the rotation axis. Black, red, and blue curves correspond to flat-ended cylinder, spherocylinder and prolate spheroid, respectively. Open symbols render the best-fitting curves. Obtained values of $q$ are also shown.}\label{comparageom}
\end{figure}

 Although Eq. \eref{Qscaspher} is only valid for spheroids, we conjecture the underlying sphere-like vs. cylinder-like interference mechanism to operate for any single nanoantenna with a similar geometry. In order to determine if there is any real substance in our guess, we present in \fref{comparageom} the calculated scattering efficiency for a single 550x30 nm Ag nanoantenna embedded in $\varepsilon_d=2.25$ assuming three different geometries: a circular cylinder with flat ends, a spherocylinder and a prolate spheroid. Numerical values are obtained from either FEM or SVM. Logarithmic scale is used for the sake of a better comparison. As can be seen, the three curves can be fitted to Eq. \eref{lineshape} even for the case of a flat-ended cylinder, although only a modest $q \approx 0.15 $ is obtained for such a high aspect ratio. Hemispherical ends result in more than a 30 percent increase of asymmetry parameter, which rises to its maximum value of $q \approx 0.92$ for prolate spheroidal geometry, where spherical and cylindrical features interact in the most efficient way. These results are consistent with our previous formal discussion on potential theory and point out the subtle balance of different contributions to light scattering.

\subsection{Spherocylinder-shaped nanorods}
\begin{figure}
 \flushright
 \includegraphics[width=0.9\textwidth]{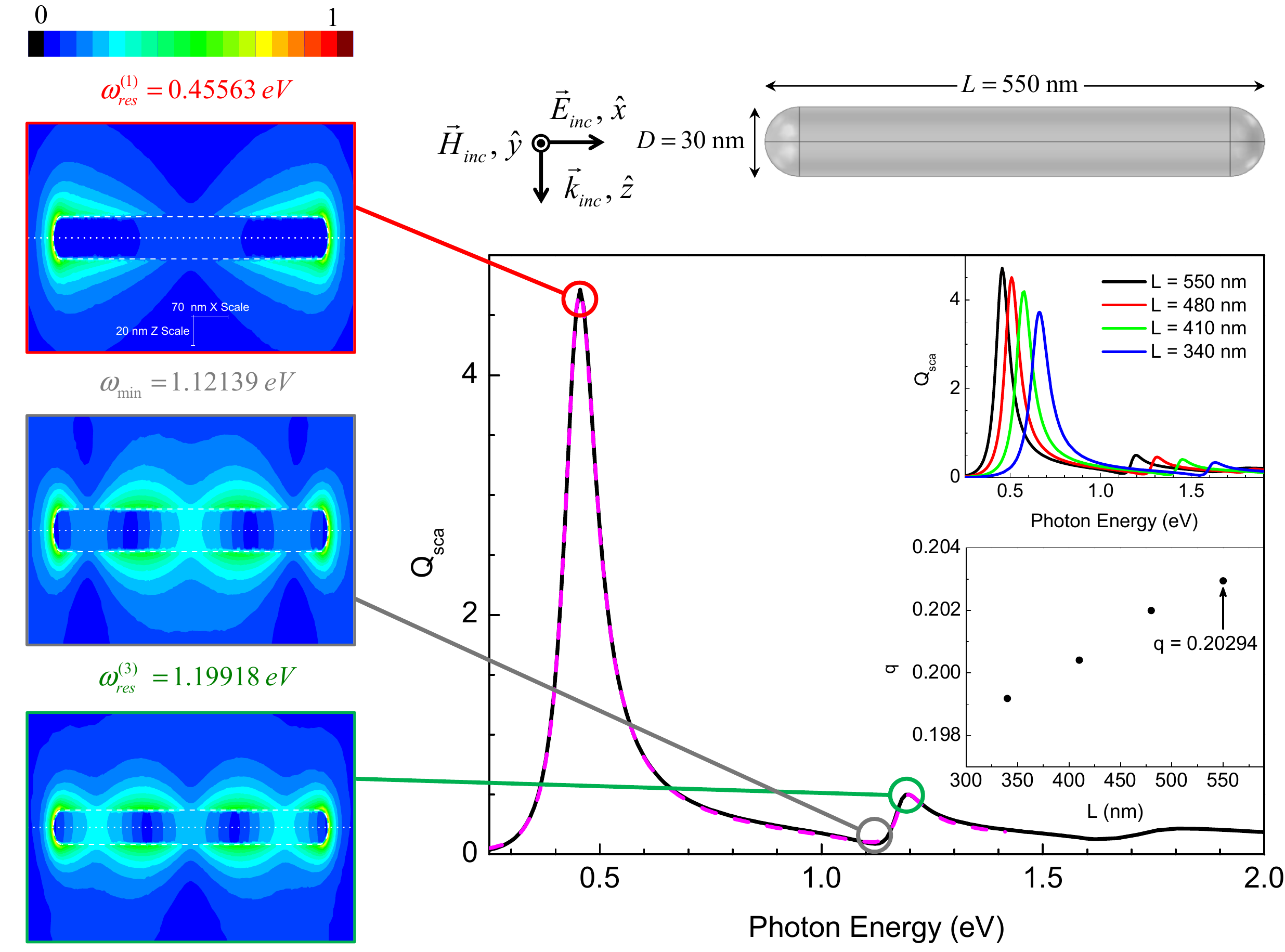}
 \includegraphics[width=0.9\textwidth]{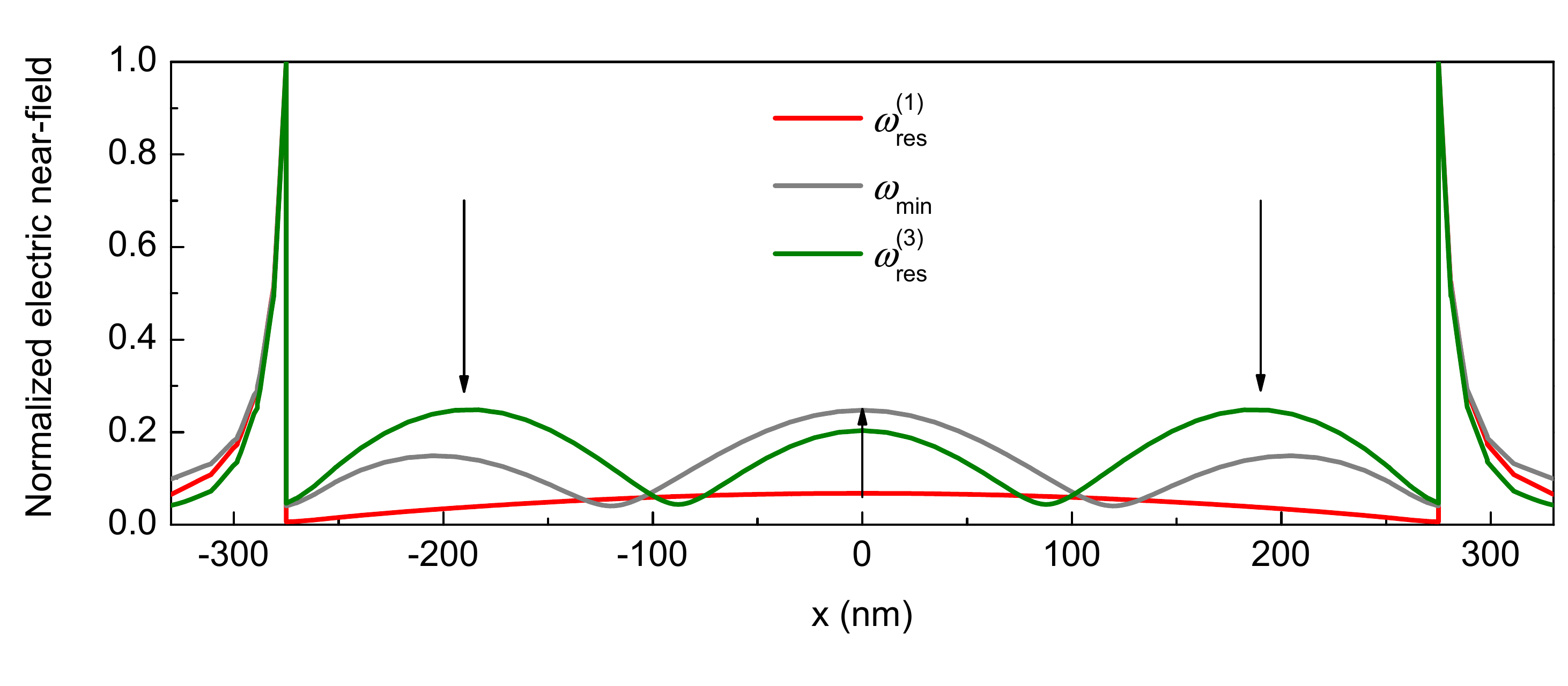}
  \caption{Calculated scattering efficiency as a function of photon energy for a single silver spherocylinder ($L=550$ nm; $D= 30$ nm) surrounded by $\varepsilon_d=2.25$. Incident field is $p$-polarized and impinges perpendicular to the long side of the rod. Dashed line renders the best-fitting curve to \ref{lineshape}. Inset panels at the right hand side show the calculated $Q_{sca}$ curves (upper) and their corresponding $q$ values (lower) for different rods with $D=30$ nm and $L$ within the $[340,550]$ nm range.The normalized electric near-field distribution in the $xz$ plane at most significant values of photon energy is presented in the three panels at the left hand side. Line profiles at $x=0$ are shown in the bottom panel.}\label{spherocyl}
\end{figure}

In \fref{spherocyl} we present a separate plot of scattering efficiency as a function of photon energy for the spherocylinder-shaped nanorod in \fref{comparageom}. As already mentioned, the asymmetric line profile can be fitted to \eref{lineshape} (dashed line). The obtained  $q=0.20294$ accounts for the moderate interaction between adjacent resonances located at $\omega_{res}^{(1)}=0.45563$ eV and $\omega_{res}^{(3)}=1.19918$ eV. Please notice that spectral features at $\omega \gtrsim 1.7 $ eV suggest the need to include subsequent resonances in \eref{lineshape2}, although that refinement is beyond the scope of our present work. Inset panels at the right hand side show the calculated $Q_{sca}$ curves (upper) and their corresponding $q$ values (lower) for different rods with $D=30$ nm and $L$ within the $[340,550]$ nm range. As expected, the interaction between adjacent resonances is significantly lower than that of prolate spheroids with the same $L,D$ parameters. (Details on fitting are presented in \ref{fitt2fano}.)

The normalized electric near-field distribution in the $xz$ plane at most significant values of photon energy is presented in the three panels at the left hand side of \fref{spherocyl}. Leaving aside the typical $n$-node quasi-standing-wave patterns at $\omega_{res}^{(1)}$ and $\omega_{res}^{(3)}$, let us concentrate on the plot for $\omega=1.12139$ eV. Given that $Q_{sca}$ reaches its local minimum within at this precise value, we expect some destructive interference to appear in spatial domain. Such interference pattern can be noticed for the field distribution inside the volume of the rod, which is limited by dashed lines.  For the sake of clarity, we present in the bottom panel the line profiles at $x=0$ (dotted lines on contour plots). In short, for $\omega=\omega_{min}$, interference cancels a significant part of the field intensity that is present at the zones marked by descending arrows for $\omega=\omega_{res}^{(3)}$. On the other hand, field intensity at the central part of the rod (ascending arrow) for $\omega=\omega_{min}$ is slightly enhanced with respect to that for $\omega=\omega_{res}^{(3)}$, thus resembling the field pattern corresponding to $\omega=\omega_{res}^{(1)}$.

\subsection{The simplest case: a rectangular nanowire}
\begin{figure}
 \flushright
 \includegraphics[width=\textwidth]{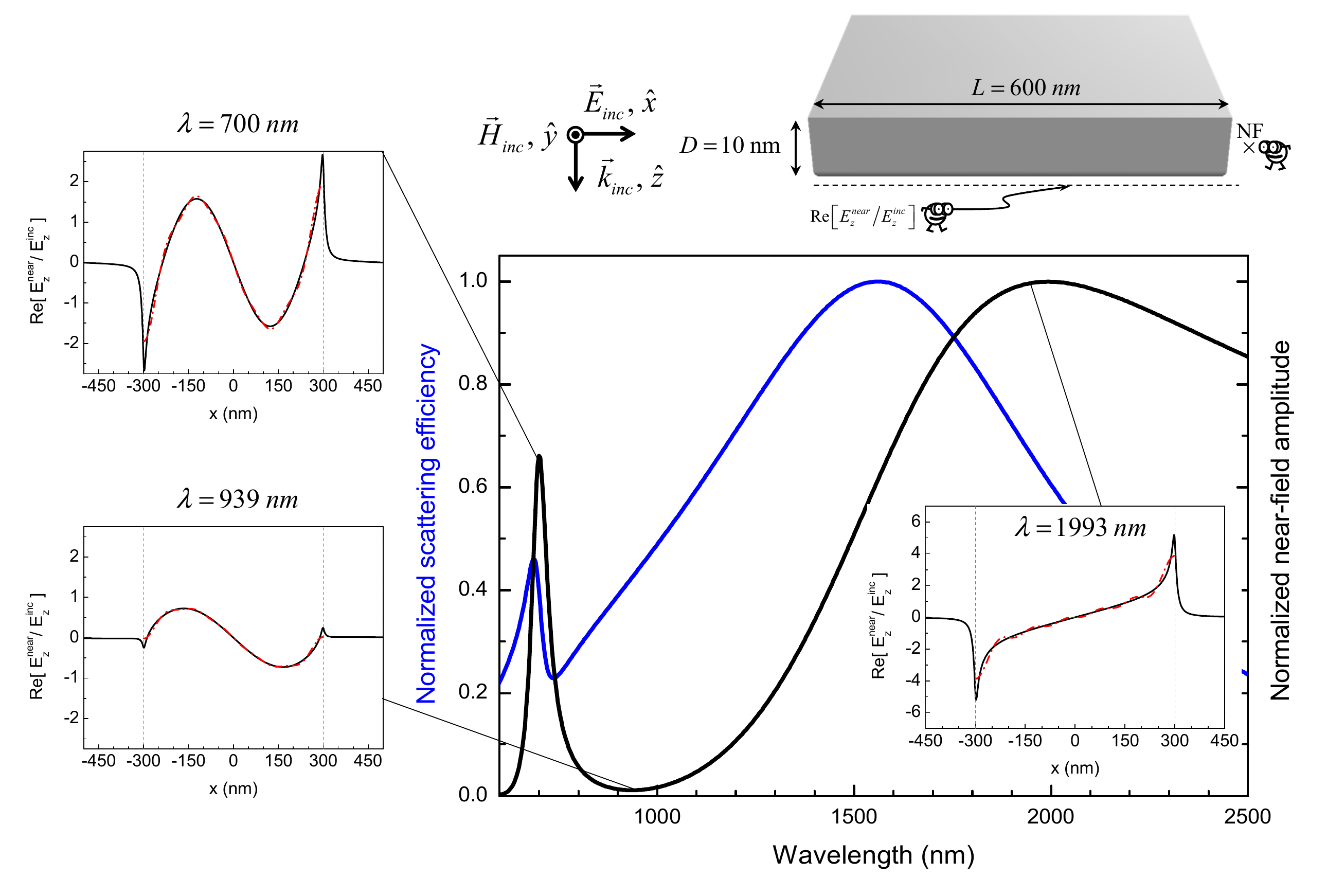}
  \caption{Calculated scattering efficiency (blue) and near-field amplitude (black) as a function of wavelength for a single Ag nanowire ($L=600$ nm, $D= 10$ nm) surrounded by $\varepsilon_d=1$. Incident field is $p$-polarized and impinges perpendicular to the $xy$ plane. The spectra have been normalized to unity. Inset panels show the calculated $\mathrm{Re}[E_z^{near}/E_z^{inc}]$ along the nanowire (solid) as well as its fitting to \ref{foursin} for $x \in [-L/2,+L/2]$ (dash-dotted), evaluated at the three most significant wavelengths in spectrum. Vertical dashed lines mark the limits of the nanowire.}\label{nanowire}
\end{figure}

For a better understanding of the elusive spatial overlap between modes, we now consider light scattering by an infinitely long rectangular nanowire with large aspect ratio, which seems to be the simplest geometry that clearly shows such behavior. In \fref{nanowire} we plot the scattering efficiency (blue curve) and the near-field amplitude (black curve) at normal incidence calculated using SIEM for a silver rectangular nanowire with $L=600$ nm and $D=10$ nm surrounded by $\varepsilon_d=1$. Near-field amplitude is now evaluated 3 nm outside the end of the nanowire and normalized to the incident field at this point. However, the two curves are then re-normalized to unity to more clearly show their spectral shift. Such a change in the position of maxima, which has recently been explained in terms of driven and damped harmonic oscillators \cite{Zuloaga2011}, will be relevant for our discussion.

In this simple geometry, the very broad dipole-like resonance in $Q_{sca}$ at $\lambda \approx 1500$ nm is more similar to the ``canonical'' Fano continuum than those in the previous configurations and therefore provides a strong spectral overlap with the narrower $3\lambda/2$-mode. With respect to spatial overlap between different longitudinal plasmon resonances, we find that it can be monitored by means of the real part of the normal component of electric field (i.e.  $E_z^{near}$ in \fref{nanowire}), which is directly proportional to surface charge distribution. Given that, in the electrostatic limit, surface charge distribution can be approximated by a sum of sinusoidal functions with argument an integer multiple of $\pi x/L$, we finally arrive to the following expression
\begin{equation}
\mathrm{Re} [E_z^{near}/E_z^{inc}](x) \approx \sum_{n=2k+1}^N A_{n} \sin \bra{[}{3} \frac{n \pi x}{ L}\bra{]}{3}\label{foursin}
\end{equation}
for which the set of coefficients $\{A_{n}\}$  can be straightforwardly determined from Fourier analysis.

According to this approximation, we expect surface charge distributions at the maxima of near-field amplitude (and not of scattering efficiency!) to have a net dipolar moment and therefore to exhibit a pattern with an odd number of nodes and charges of opposite sign at its ends. Surface charge distribution will gradually modify as wavelength varies and adjacent resonances come into play. In fact, somewhere within the intermediate region, destructive interference between modes reduces near-field amplitude to a minimum, as can be seen in \fref{nanowire}. In order to quantify such features we find it useful to define the following auxiliary magnitudes: as a measure of the contribution of $n$-resonance at a given wavelength, we introduce its relative weight as
\begin{equation}
f_{n} \equiv \frac{\displaystyle A_{n}^2}{\displaystyle \sum_{m}A_{m}^2}\label{fval}
\end{equation}
which will vary from $f_{n} \approx 1$ in the vicinity of $n$-maximum to $f_{n} \approx 0$ as the charge reaches its subsequent resonance. Additionally, we take the right hand member of \eref{foursin} as the measure of surface charge at $x=+L/2$:
\begin{equation}
 \Delta_{\sigma} \equiv\sum_{n=2k+1}^N (-1)^{(n-1)/2} A_n \approx \mathrm{Re} [E_z^{near}/E_z^{inc}]_{x= +L/2}\label{Delta}
\end{equation}

Inset panels in \fref{nanowire} show the calculated $\mathrm{Re}[E_z^{near}/E_z^{inc}]$ along the nanowire (solid) as well as its Fourier series up to $N=11$ for $x \in [-L/2,+L/2]$ (dash-dotted), evaluated at the three most significant wavelengths in spectrum: At $\lambda=1993 $ nm, surface charge exhibits a typical dipole-like distribution that is in good agreement with the obtained $f_1= 0.8304,f_3=0.08887$, whereas $f_{\Sigma}\equiv \sum_{n>3}f_{n}=0.08075$ accounts for the rest of contributions. A similar behavior can be found at $\lambda=700 $ nm, but longitudinal mode with $n=3$ now plays the main role and the values of $f_1,f_3$ are almost interchanged, namely $0.0785,0.8279$. For the relative minimum of near-field amplitude at $\lambda=939$ nm, the situation is completely different. Although obtained values of $f_1, f_3,f_{\Sigma}$  are not too different of those at $\lambda=1993$ nm (see \tref{tbl:fval}), the field profile resembles that of $\lambda=700$ nm. However, surface charge reduces drastically at the ends of the nanowire, which is consistent with the expected minimum in dipolar moment. In fact, the obtained $\Delta_{\sigma}=0.01947$ is two orders of magnitude less than those for $\lambda=1993, 700$ nm (see last row in \ref{tbl:fval}) and nearly approaches to the condition of complete destructive interference at $x= \pm L/2$ in \eref{foursin}, which is given by $\Delta_{\sigma}=0$. We find this direct evidence of spatial overlap between interfering plasmon modes to confirm the consistency of our Fano-like model when applied to a generic nanorod geometry.
\begin{table}
  \caption{\label{tbl:fval}Obtained values of $f_1,f_3,f_{\Sigma},\Delta_{\sigma}$ for inset panels in \fref{nanowire}.}
  \begin{indented}
  \item[]\begin{tabular}{@{}llll}
    \br
    & $\lambda=1993$ nm  & $\lambda=939$ nm & $\lambda=700$ nm \\
    \mr
    $f_1$ & 0.8304  &  0.7289  & 0.0785\\
    $f_3$ & 0.08887 &  0.2196 & 0.8279\\
    $f_{\Sigma}$ & 0.08075 & 0.05148	& 0.09362 \\
    $ \Delta_{\sigma}$ & 3.8737 &  0.01947  & 1.95\\
    \br
  \end{tabular}
  \end{indented}
\end{table}

\section{Conclusions}
In conclusion, the emergence of asymmetric line profiles in the total scattering spectra of single metallic nanorods acting as half-wave nanoantennas is explained in terms of Fano-like interference between adjacent, odd-order plasmon resonances. Such a feature can be understood as originated from the interplay between the different contributions that properly describe the non-axisymmetric part of the scattered field. Our analytical and numerical results show that Fano resonances can be excited on diverse geometries (elongated nanospheroids, nanorods and nanowires), provided that interacting modes overlap not only in energy but also spatially (non-orthogonality). This finding makes single-particle nanoantennas especially suitable for a wealth of applications where the sharp, environment-sensitive Fano-like spectral profile can be crucial \cite{Yanik2011}. Moreover, the underlying mechanism of the resulting Fano resonances has far reaching implications in exploring other single nanoparticle configurations designed for specific applications.

\ack The research presented in this paper is supported by the Spanish ``Ministerio de Ciencia e Innovaci\'on'' (projects Consolider-Ingenio EMET CSD2008-00066 and NANOPLAS FIS2009-11264) and the ``Comunidad de Madrid'' (MICROSERES network S2009/TIC-1476). R. Paniagua-Dom\'{\i}nguez acknowledges support from CSIC through a JAE-Pre grant. The authors also acknowledge Prof. N. V. Voshchinnikov for kindly providing an updated version of SVM code and some practical indications about its usage.

\appendix

\section{Calculation techniques}
\label{calcu}
Calculated scattering efficiencies for spheroids were attained by means of a modified version of the F77-code made publicly available at the web site of the Jena-St.Petersburg Database of Optical Constants (JPDOC) \cite{JPDOC}. Detailed description of SVM formalism and benchmark results for the code can be found in References \cite{Voshchinnikov1993} and \cite{Voshchinnikov2000}, respectively.

Results for cylindrically-shaped rods with either flat or hemispherical ends were obtained with the RF module of COMSOL Multiphysics finite element software, version 4.2. The computational domain consisted on the nanorod  (which, for the case of the spherocylinder, was modeled as the union of a cylinder, with length $L-D$ and diameter $D$, and two spheres of radii $R=D/2$ centered in each of the cylinder flat faces) and two concentric spheres of radii $R_1=0.7(L-D)$ and $R_2=1.1(L-D)$ surrounding it. The space between the smallest sphere and the rod models the medium in which the nanoantenna is embedded (glass). The subdomain comprised between the two spheres is set to be a Perfectly Matched Layer (PML), which absorbs all scattered fields. The mesh was constructed with the software built-in algorithm, which generates a free mesh consisting on tetrahedral elements. The subdomain representing the rod was constructed in all cases with mesh elements constrained to a maximum element edge size of 15~nm and a growth rate of 1.4, meaning that adjacent elements to a given one should not be bigger than 1.4 times the size of it. This allows to correctly map the near-field zone where the fields decay rapidly. The rod is adaptively meshed with elements that are finer in the cups of the rod and coarser along the cylindrical body. As an illustrative example, in the case of a 340x30~nm rod, the subdomain representing the rod consisted on 18337 mesh elements, while the subdomains modeling the external medium and the PML consisted on 48783 mesh elements. The system was solved using PARDISO direct solver and involved, in this case, 427502 degrees of freedom. As the length of the rod was increased the number of mesh elements grew, and so did the number of degrees of freedom to be solved. The scattering efficiency was calculated as
\begin{equation}
Q_{sca}=\frac{1}{A_{rod} \sqrt{\epsilon_{d}} |\mathbf{E_{inc}}|^2 R_{calc}^2} \int{|\mathbf{E_{far}}|^2 R_{calc}^2 d\Omega}.
\end{equation}
The integration was performed over a spherical boundary of radius $R_{calc}$. We chose it to be the boundary between the embedding medium and the PML, thus $R_{calc}=R_{1}$. $A_{rod}$ is the transverse geometrical area of the rod, which, in the case of normal incidence, is simply $2RL+(\pi-4)R^2$. $\mathbf{E_{inc}}$ is the incident field amplitude, and $\mathbf{E_{far}}$ is the vectorial far-field of the scattered field obtained with the COMSOL implementation of the Stratton-Chu formula \cite{Stratton1939}. Please notice that, as previously pointed out by other authors \cite{Knight2008}, COMSOL automatically introduces the factor $R_{calc}^2$ when performing the integration, so that it has to be canceled in order to obtain the correct value for $Q_{sca}$.

For calculations of light scattering by an infinitely long nanowire in \ref{nanowire}, we made use of our own implementation \cite{Giannini2007,Roger2011} of Green's theorem surface integral equations, which are written for parametric surfaces describing particles with arbitrary shape by means of Gielis' ``Superformula'' \cite{Gielis2003}.

The dielectric function of Ag was modeled as a sum of Drude and Lorentz terms \cite{Rodrigo2008} for all geometries.

\section{Details on fitting of scattering efficiency to Fano-like lineshape}
\label{fitt2fano}

\begin{figure}
  \flushright
  \includegraphics[width=0.82\textwidth]{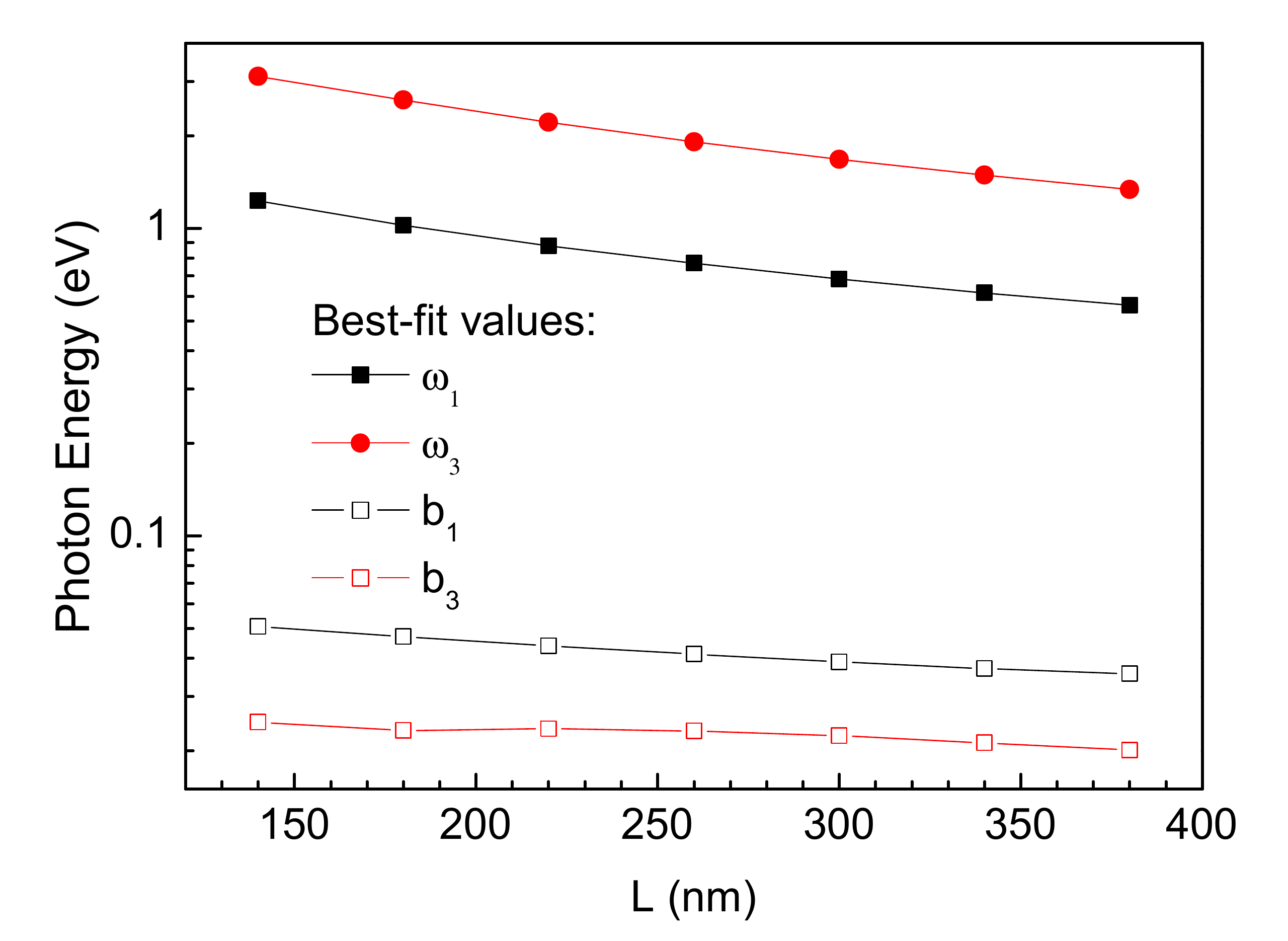}\\
  \caption{Obtained values of $\omega_1,\omega_3,b_1,b_3$ for every $L$ in \fref{Qscavsvlan}.}\label{fittedval1}
\end{figure}
With respect to the slowly varying amplitude $A(\omega)$ in \eref{lineshape2}, we have heuristically chosen
\begin{equation}
A(\omega)=\frac{A_0}{(1+\omega-\omega_{min})^\beta}
\end{equation}
where $\omega_{min}=0.5\omega_1$, $\beta \lesssim 1$ and $A_0 \equiv |A_0|e^{i\varphi'}$. Best-fitting curves of $Q_{sca}$ to $|f(\omega)|^2$ presented in \fref{QscavseV}, \fref{obliki} and \fref{spherocyl}  were obtained by means of an iterative implementation of Levenberg-Marquardt algorithm \cite{Levenberg1944, Marquardt1963}.

In order to place in context the information on the evolution of $q$ parameter included in \fref{QscavseV}, we summarize in \fref{fittedval1} all our fitting duties with respect to $\omega_1,\omega_3,b_1,b_3$ for all spheroids in \fref{Qscavsvlan} ($L \in [100,400]$ nm, $D=30$ nm, $\varepsilon_d=2.25$). Logarithmic scale is used for the sake of readability. As expected from standard antenna theory, $\omega_1, \omega_3$ scale linearly with $1/L$. Interestingly, $b_1/b_3$ ratio slightly decreases as $L$ increases, thus fading the difference between ``continuum-like'' and ``localized-like'' resonances.

\begin{figure}
  \flushright
  \includegraphics[width=0.82\textwidth]{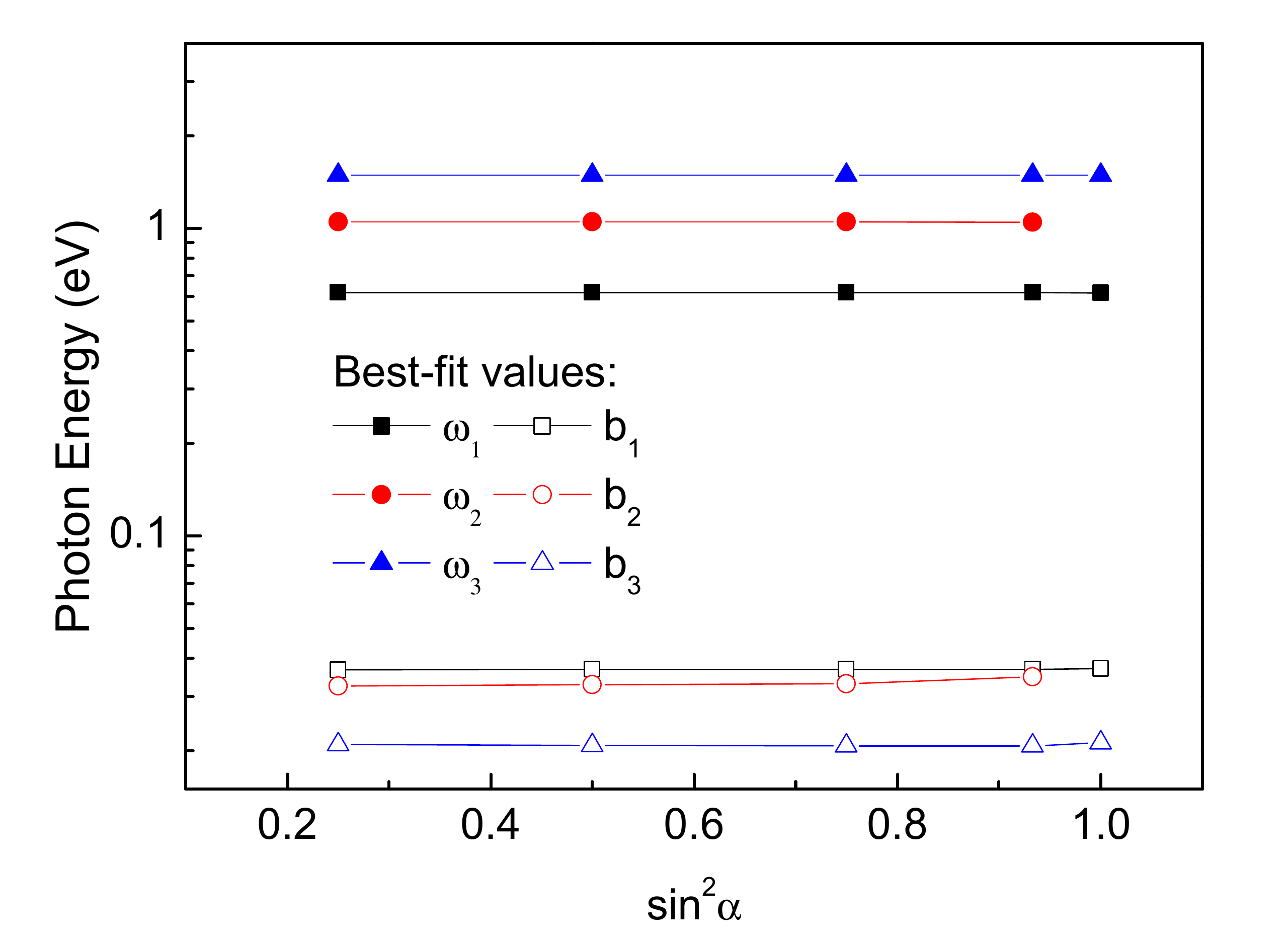}\\
  \caption{Obtained values of $\omega_1,\omega_2,\omega_3,b_1,b_2,b_3$ for every angle $\alpha$ in \fref{obliki}.}\label{fittedval2}
\end{figure}

For the case of $p$-polarized light impinging with oblique incidence on a single Ag spheroid surrounded by $\varepsilon_d=2.25$ (see \fref{obliki}),   parameters $\omega_1,\omega_2,\omega_3,b_1,b_2,b_3$  remain almost invariant with respect to angle $\alpha$, as can be seen in \ref{fittedval2}. This is consistent with the lack of interaction between resonances with different parity.

\begin{figure}
  \flushright
  \includegraphics[width=0.82\textwidth]{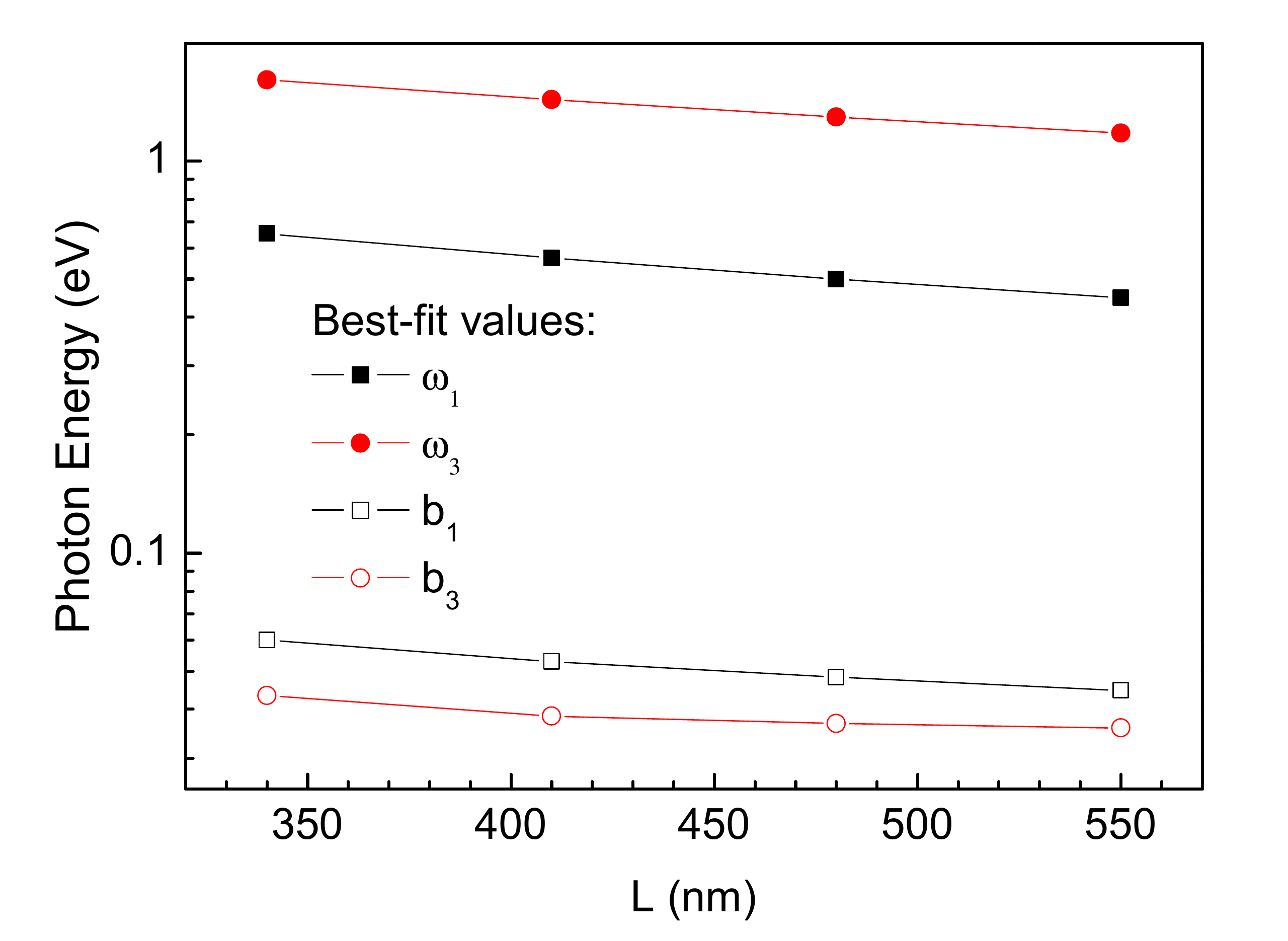}\\
  \caption{Obtained values of $\omega_1,\omega_3,b_1,b_3$ for every $L$ in \fref{spherocyl}.}\label{fittedval3}
\end{figure}

Finally, we present in \fref{fittedval3} the obtained $\omega_1,\omega_3,b_1,b_3$ for all spherocylinders in \fref{spherocyl} ($L \in [300,600]$ nm, $D=30$ nm, $\varepsilon_d=2.25$). Except for numerical values, all parameters $L$ follow the same trends than in \ref{fittedval1}, thus confirming the validity of our Fano-like model when applied to a generic nanorod geometry.

\section*{References}


\end{document}